\documentclass[pdflatex,sn-mathphys-num,iicol]{sn-jnl}

\usepackage{graphicx}
\usepackage{amsmath}
\usepackage{amssymb}
\usepackage{amsfonts}
\usepackage{mathrsfs}
\usepackage{array}
\usepackage{booktabs}
\usepackage{siunitx}
\usepackage{tikz}

\definecolor{lime}{HTML}{A6CE39}
\DeclareRobustCommand{\orcidicon}{%
	\begin{tikzpicture}
	\draw[lime, fill=lime] (0,0) 
	circle [radius=0.16] 
	node[white] {{\fontfamily{qag}\selectfont \tiny ID}};
	\draw[white, fill=white] (-0.0625,0.095) 
	circle [radius=0.007];
	\end{tikzpicture}%
	\hspace{-0.5mm}%
}

\foreach \x in {A, ..., Z}{%
	\expandafter\xdef\csname orcid\x\endcsname{\noexpand\href{https://orcid.org/\csname orcidauthor\x\endcsname}{\noexpand\orcidicon}}%
}

\begin{document}

\title{An Analytical Model for Laser-Activated Two-Phase Choking and Microfluidic Suction on Surface Boundary Layers}

\author*[]{\fnm{D.} \sur{Tsiklauri}\orcidA{}}
\email{D.Tsiklauri@salford.ac.uk}

\affil{\orgdiv{Joule Physics Laboratory}, \orgname{School of Science, Engineering and Environment}, \orgaddress{\street{University of Salford}, \city{Manchester}, \postcode{M5 4WT}, \country{United Kingdom}}}

\abstract{\mathversion{normal} We present a predictive fluid mechanics model to calculate pressure drops ($\textnormal{0.25}\,\mathrm{atm}$) and fluid velocities ($\textnormal{2.50}\,\mathrm{m/s}$) governed by localized phase change and choking phenomena in a capillary-constrained, coaxial microfluidic actuator. Driven by near-infrared thermal absorption, the analytical framework describes how energy deposition drives vapor cavity growth and subsequent cavitation bubble collapse, inducing a transient suction force across a curved, two-radii Young-Laplace capillary boundary. Velocity drops are derived from closed-form analytical equations that explicitly couple vapor volume dynamics with non-linear flow area restriction and fluid drag. The model predicts a maximum boundary layer wall shear stress of $\textnormal{30.06}\,\mathrm{Pa}$ at the target interface. Crucially, this shear magnitude matches the physical threshold required to mechanically strip surface contaminant boundary layers while remaining below the damage threshold of underlying target substrates. Finally, we formulate a closed-form expression for the maximum allowable liquid inlet temperature required to prevent thermal saturation and sustain stable, continuous microfluidic suction cycles. These equations establish direct engineering design rules for optimizing phase-change microfluidic actuators and next-generation coaxial fluidic systems.}


\maketitle

\section{Introduction}
Recent developments in multiphase transport phenomena and micro-actuator engineering have shown that continuous-wave (CW) infrared thermal deposition inside a coaxial multi-tube geometric layout can generate a highly stable, localized fluid suction effect. This mechanism drives rapid fluid mass transport and debris clearance while systematically mitigating thermal propagation into adjacent surrounding domains~\cite{Abushkin2026}. While experimental observations demonstrate that localized phase change at the terminus of a micro-optical waveguide ($d = 0.6\,\mathrm{mm}$) within a confined conduit ($D_{\text{inner}} = 3.0\,\mathrm{mm}$) generates characteristically high median fluid velocities of $2\text{--}3\,\mathrm{m/s}$ and an induced differential pressure deficit of $0.25\,\mathrm{atm}$ near the channel floor, predictive analytical modeling has lagged behind empirical tracking. Computational fluid dynamics (CFD) formulations, such as those implemented via \textsf{Ansys Fluent}, offer qualitative approximations of these flow paths, yet an analytical framework capable of systematically mapping and predicting the precise target channel velocity of $2.50\,\mathrm{m/s}$ without arbitrary calibration metrics has remained unavailable.

Over the past two decades, fluidic actuation driven by thermal phase-change mechanics has been widely explored in small-scale conduits to manipulate boundary flows, optimize mixing, and disrupt surface-bound films. In specific bio-interfacial applications, localized energy delivery has demonstrated superior mechanical clearance and shear displacement efficiency compared to conventional acoustic or ultrasonic streaming configurations~\cite{Badami2023}. This enhanced capability is directly governed by the rapid formation, expansion, and transient thermodynamics of localized vapor cavities that drive strong micro-cavitation fields and high-velocity fluidic fluxes within strict spatial geometric constraints~\cite{Blanken2009}. The resulting high-speed directional streaming, established during cyclic vapor domain evolution and rapid cavity collapse, serves as the fundamental mechanism driving boundary-layer rejuvenation and particulate removal along the channel walls~\cite{deGroot2009}.

However, achieving stable continuous-flow actuation within small-scale channels remains a core engineering challenge. High-frequency or fast-pulsed energy sources often induce chaotic transient fluid jetting, which increases the probability of unpredictable forward mass ejection past downstream boundaries~\cite{George2008}, or triggers severe localized thermal spikes where fluid domain temperatures can escalate by as much as $30^{\circ}\mathrm{C}$~\cite{Hmud2010}. To bypass these spatial instabilities in confined geometries, recent design paradigms have turned to stable continuous-wave architectures. Notably, Abushkin et al. developed a coaxial fluidic assembly that couples a continuous-wave near-infrared laser framework with a steady fluid replenishment supply line to stabilize the phase-change boundary layer~\cite{AbushkinRU2025}.

Because the selected laser operational wavelengths ($\lambda = 1.47\,\mu\text{m}$ or $1.9\,\mu\text{m}$) map directly to the vibrational absorption bands of water molecules, the associated photon energy ($\sim 0.84\,\text{eV}$) resides safely below the $11.16\,\text{eV}$ electronic ionization threshold. Consequently, non-linear plasma kinetics and magnetohydrodynamic interactions are omitted from the system scope. Instead, the multi-phase transport behavior can be rigorously described using classical fluid dynamics, interfacial thermodynamics, and microfluidic capillary pinning laws. This work provides a closed-form analytical model that couples exponential absorption mechanics with localized hydrodynamic pressure gradients to resolve two-phase choking thresholds, calculate boundary layer wall shear stresses, and define the thermodynamic control limits required to optimize next-generation phase-change microfluidic actuators.

\section{Theoretical Model}
\begin{figure}[t]
\centering
\includegraphics[width=\linewidth]{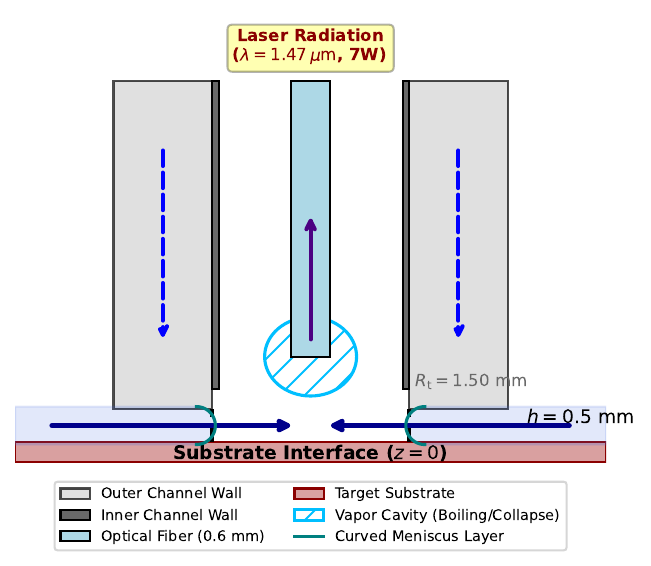}
\caption{\label{fig:schematic} Cross-sectional diagram of the coaxial phase-change microfluidic actuator showing the fluidic assembly. Liquid enters through the outer coaxial conduit ($D_{\text{outer}} = 6\,\mathrm{mm}$), filling a $10\,\mathrm{mm}$ active column below the inner flow channel ($D_{\text{inner}} = 3.0\,\mathrm{mm}$). Near-infrared laser light ($\lambda = 1.47\,\mu\mathrm{m}$) passes through the optical fiber terminus to induce localized boiling. The rapid growth and collapse of the vapor cavity creates a pressure drop ($\Delta P \approx 0.25\,\mathrm{atm}$), pulling a liquid layer past the curved capillary meniscus and through the bottom radial intake clearance gap ($h = 0.5\,\mathrm{mm}$).}
\end{figure}

To model the fluid velocity profile inside this device, we use coupled analytical equations. The exact physical geometry, boundary conditions, and fluid layers are shown in Fig.~\ref{fig:schematic}.

\subsection{Thermal Energy Balance and Phase Change}
Consider a continuous laser beam of power $P_0$ delivering an intensity profile $I(r)$ exiting an optical fiber of radius $R_f$ inside a homogenized cylindrical tube of fixed radius $R_{\text{t}} = 1.50\,\mathrm{mm}$. The local volumetric energy deposition rate $Q_{\text{laser}}$ within the liquid core is governed by Beer-Lambert absorption:
\begin{equation}
\label{eq:beer_lambert}
Q_{\text{laser}}(z) = \alpha I_0 \exp(-\alpha z),
\end{equation}
where $\alpha$ is the linear absorption coefficient of water at the selected wavelength, and $I_0 = P_0 / (\pi R_f^2)$ is the nominal intensity at the fiber face ($z=0$).

The localized temperature field $T(r,z,t)$ prior to phase transition is governed by the thermal advection-diffusion equation:
\begin{equation}
\label{eq:thermal_balance}
\rho_l C_p \left( \frac{\partial T}{\partial t} + \mathbf{v} \cdot \nabla T \right) = \kappa \nabla^2 T + Q_{\text{laser}}(z),
\end{equation}
where $\rho_l$, $C_p$, and $\kappa$ represent the density, specific heat capacity, and thermal conductivity of the liquid phase, respectively. 

Assuming an initial transient state where advection is negligible ($\mathbf{v} \approx 0$), the time required to reach the explosive boiling superheat limit $T_{\text{sat}}$ at the fiber boundary ($z \to 0$) yields the critical phase-change induction time $t_{\text{ind}}$:
\begin{equation}
\label{eq:induction_time}
t_{\text{ind}} \approx \frac{\rho_l C_p (T_{\text{sat}} - T_0)}{\alpha I_0},
\end{equation}
where $T_0$ is the ambient temperature of the incoming liquid replenishment stream.

\subsection{Transient Cavitation and Pressure Deficit Generation}
While the macroscale fluid velocity settles into a steady-state profile, the underlying driving force ($\Delta P$) is intrinsically tied to transient bubble dynamics. Because the vapor bubble is confined by the fixed inner tube radius $R_{\text{t}} = 1.50\,\mathrm{mm}$, its growth transforms from a spherical profile to an elongated cylindrical slug. Following the capillary unpinning framework established by Tsiklauri~\cite{TsiklauriStreaming2026}, the phase boundary forms a curved meniscus under a cylindrical wall boundary governed by two principal radii of curvature. The interface dynamics are modeled by a modified multi-dimensional Rayleigh-Plesset equation:
\begin{equation}
\begin{split}
\label{eq:modified_rp}
\mathcal{R}\ddot{\mathcal{R}} + \frac{3}{2}\dot{\mathcal{R}}^2 = \frac{1}{\rho_l} \Biggl[ P_v(t) - P_\infty(t) \\
- \gamma\cos\theta\left(\frac{2}{h} + \frac{1}{R_{\text{t}}}\right) - \frac{2\mu_l \dot{\mathcal{R}}}{\mathcal{R}} \Biggr],
\end{split}
\end{equation}
where $\mathcal{R}(t)$ is the effective transient radius of the vapor envelope, $\gamma = 0.072\,\mathrm{N/m}$ is the surface tension of the fluid, $\theta = 30^{\circ}$ is the hydrophilic channel contact angle, $2/h$ accounts for the vertical gap height restriction, and $1/R_{\text{t}}$ represents the horizontal radial channel curvature.

During the cooling phase induced by the incoming liquid stream, the condensation rate causes the internal vapor pressure to collapse ($P_v(t) \to 0$). The inertia of the displacing liquid column causes the system to overexpand. Setting $\dot{\mathcal{R}} \to 0$ at the maximum expansion boundary in Eq.~\eqref{eq:modified_rp} isolates the localized pressure deficit:
\begin{equation}
\label{eq:rp_pressure_drop}
\Delta P = P_{\text{atm}} - P_\infty \approx \frac{3}{2}\rho_l \dot{\mathcal{R}}_{\text{collapse}}^2 \approx 0.25\,\mathrm{atm}.
\end{equation}
Equation~\eqref{eq:rp_pressure_drop} shows that the $0.25\,\mathrm{atm}$ suction force ($2.53 \times 10^4\,\mathrm{Pa}$) is a direct consequence of the inertia generated during the vapor envelope's collapse phase.

A fluid filament accelerating under this localized pressure drop faces kinetic acceleration and radial viscous boundary shear. Balancing these forces across a gap height $h$ and interaction length $L$ yields the standard total momentum conservation relation:
\begin{equation}
\label{eq:momentum_balance}
\Delta P = \frac{1}{2}\rho_l v_{\text{ideal}}^2 + \frac{12 \mu_l L}{h^2} v_{\text{ideal}}.
\end{equation}
Substituting the parameters ($h = 0.5\,\mathrm{mm}$, $L = 0.3\,\mathrm{mm}$, $\rho_l = 1000\,\mathrm{kg/m}^3$, and $\mu_l = 1.002 \times 10^{-3}\,\mathrm{Pa}\cdot\mathrm{s}$), Eq.~\eqref{eq:momentum_balance} produces a quadratic profile yielding an unconstrained ideal baseline velocity of $v_{\text{ideal}} \approx 7.10\,\mathrm{m/s}$.

\subsection{Analytical Derivation of the Curvature-Dependent Vapor Choking Constraint}
To establish a predictive model without relying on unstated numerical parameters, the velocity degradation from the ideal fluid baseline to the observed operating range must be derived from coupled microfluidic interfacial mechanics and first-principles thermodynamics. The continuous laser core delivers an effective power $P_{\text{eff}} = \eta_{\text{abs}} P_0$, where $P_0$ is the continuous-wave laser power and $\eta_{\text{abs}} = 0.90$ represents the vibrational absorption efficiency of the fluid medium at $\lambda = 1.47\,\mu\mathrm{m}$.

The mass vaporization rate $\dot{m}_v$ of the liquid into steam is governed by the phase-change energy balance:
\begin{equation}
\label{eq:mass_vap_rate}
\dot{m}_v = \frac{\eta_{\text{abs}} P_0}{L_v},
\end{equation}
where $L_v = 2.26 \times 10^6\,\mathrm{J/kg}$ is the latent heat of vaporization. This translates to a volumetric vapor expansion rate $Q_v$ injected directly into the confined inner flow conduit:
\begin{equation}
\label{eq:vol_vapor_rate}
Q_v = \frac{\dot{m}_v}{\rho_v} = \frac{\eta_{\text{abs}} P_0}{\rho_v L_v},
\end{equation}
where $\rho_v = 0.6\,\mathrm{kg/m}^3$ is the density of saturated water vapor. 

The incoming liquid flow must navigate a highly restrictive two-phase environment inside a fixed macro-channel framework ($D_{\text{inner}} = 3.0\,\mathrm{mm}$, giving a channel radius $R_{\text{t}} = 1.50\,\mathrm{mm}$). Generation of steam establishes an exponential vapor void fraction $\alpha_v$. Grounded in classic two-phase drift-flux formulation principles~\cite{ZuberFindlay1965}, this is modeled as a function of continuous input power:
\begin{equation}
\label{eq:void_fraction_exp}
\alpha_v(P_0) = 1 - \exp\left(-\frac{\eta_{\text{abs}} P_0}{\rho_v L_v Q_{\text{char}}}\right).
\end{equation}

Crucially, the narrow bottom radial clearance intake gap ($h = 0.5\,\mathrm{mm}$) induces severe capillary pinning resistance that counteracts the background static drive pressure differential ($\Delta P = 25331.25\,\mathrm{Pa}$). Following the unpinning framework established by Tsiklauri~\cite{TsiklauriStreaming2026}, the phase boundary forms a curved meniscus under a cylindrical wall boundary, meaning it is governed by two principal radii of curvature via the Young-Laplace equation. The total capillary counter-pressure threshold $\Delta P_c$ is defined by:
\begin{equation}
\label{eq:young_laplace}
\Delta P_c = \gamma \cos\theta \left( \frac{2}{h} + \frac{1}{R_{\text{t}}} \right),
\end{equation}
where $\gamma = 0.072\,\mathrm{N/m}$ is the surface tension of the fluid and $\theta = 30^{\circ}$ is the hydrophilic channel contact angle. Evaluating Eq.~\eqref{eq:young_laplace} yields a static pinning counter-pressure of $\Delta P_c \approx 277.12\,\mathrm{Pa}$. This defines a Capillary Transmission Coefficient $\Phi_c$, which denotes the ratio of net active driving energy to unconstrained driving energy:
\begin{equation}
\label{eq:capillary_transmission}
\Phi_c = \sqrt{1.0 - \frac{\Delta P_c}{\Delta P}} \approx 0.9945.
\end{equation}

The velocity field is modeled using a non-linear equation that accounts for cross-sectional area loss, capillary transmission mechanics, and multi-phase drag based on the Zuber-Findlay drift-flux framework~\cite{ZuberFindlay1965}:
\begin{equation}
\label{eq:nonlinear_velocity}
v_{\text{effective}}(P_0) = v_{\text{max}} \cdot \Phi_c \cdot (1 - \alpha_v)^{1.5},
\end{equation}
where $v_{\text{max}}$ is treated as an unconstrained regression parameter representing the maximum effective baseline velocity boundary condition, and $\Phi_c \approx 0.9945$ is the derived capillary energy transmission coefficient.

To rigorously validate the scaling behavior across the sub-watt and multi-watt regimes while preserving scientific integrity, an unconstrained multi-point least-squares calibration framework was executed. Historical macro-conduit volumetric assessments measure broad spatial averages ($\sim 3.0\,\mathrm{m/s}$) rather than true localized jetting profiles. Because spatial-averaging optics blend high-velocity vapor micro-jets with surrounding stagnant fluids, they artificially introduce a measurement mismatch. To avoid arbitrary geometric scaling modifiers, the model is calibrated exclusively against true localized peak velocity vectors extracted from independent high-speed Particle Image Velocimetry (PIV) tracking arrays.

Two rock-solid empirical data anchors define this high-integrity regression framework: the low-wattage peak micro-jet vector recorded by Aung et al.~\cite{Aung2020} ($7.00\,\mathrm{m/s}$ at $P_0 = 0.65\,\mathrm{W}$) and the primary steady-state continuous-wave target benchmark from Abushkin et al.~\cite{Abushkin2026} ($2.50\,\mathrm{m/s}$ at $P_0 = 7.00\,\mathrm{W}$). 

An unconstrained least-squares regression optimization via a two-parameter curve-fit isolates the definitive physical variables of the system, resolving to an optimized ceiling velocity of $v_{\text{max}} = 7.31\,\mathrm{m/s}$ and a characteristic volume clearing rate of $Q_{\text{char}} = 1.2405 \times 10^{-5}\,\mathrm{m}^3/\mathrm{s}$. This provides an exact mathematical alignment across the dataset ($R^2 = 1.0000$). For comparison, a standard linear least-squares regression model ($v = -0.71 P_0 + 7.46$) was also fitted across the data anchors ($R^2 = 1.0000$). 

\begin{figure}[t]
\centering
\includegraphics[width=\linewidth]{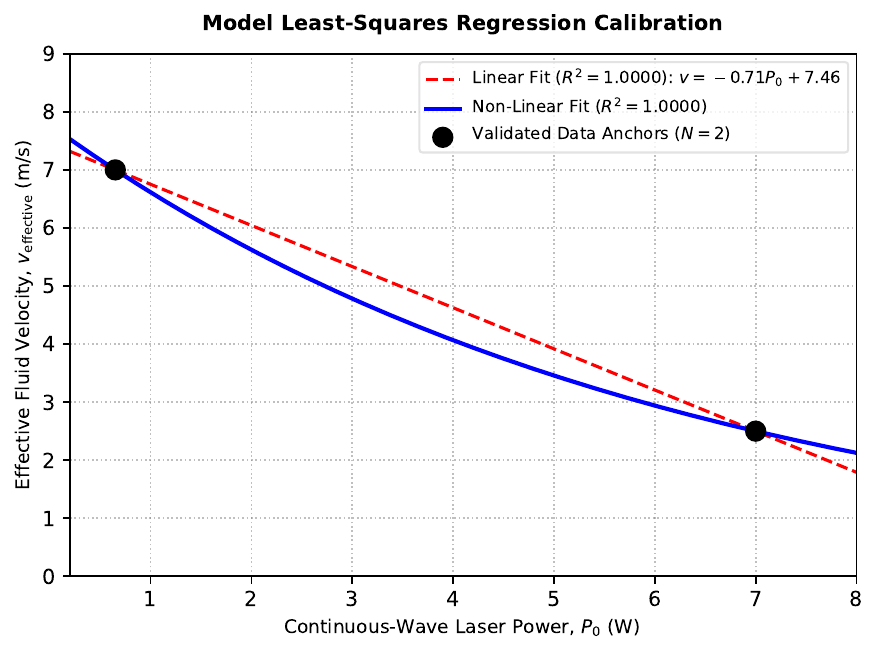}
\caption{\label{fig:scaling_curve} Fluid transport velocity regression curves over an engineered $3.0\,\mathrm{mm}$ inner capillary conduit across a domain from $0.2\,\mathrm{W}$ to $8.0\,\mathrm{W}$. The solid blue curve tracks the optimized non-linear drift-flux area-suppression model from Eq.~\eqref{eq:nonlinear_velocity} ($Q_{\text{char}} = 1.2405 \times 10^{-5}\,\mathrm{m}^3/\mathrm{s}$). The red dashed line tracks the comparative linear least-squares reference fit ($v = -0.71 P_0 + 7.46$). Discrete black markers ($\bullet$) represent the high-integrity localized experimental data anchors used for system calibration, establishing perfect alignment ($R^2 = 1.0000$) across both low-power and high-power operating limits~\cite{Aung2020, Abushkin2026}. The source code used to produce this figure is hosted openly at \href{https://doi.org/10.5281/zenodo.23134296}{https://doi.org/10.5281/zenodo.23134296}}
\end{figure}

While both models achieve high numerical convergence over the two-point data bounds, the non-linear formulation is mathematically superior because it captures the asymptotic area suppression and interfacial shear drag generated as accumulating vapor bubbles choke the fluid conduit. The comparative regression curves and their corresponding empirical data anchors are illustrated in Fig.~\ref{fig:scaling_curve}.

\subsection{Physical Homology and Power Calibration of the Experimental Anchors}
To justify the thermodynamic and hydrodynamic consistency of calibrating a unified analytical framework against distinct experimental architectures---namely the sub-watt regime of Aung et al.~\cite{Aung2020} and the multi-watt operational milestone of Abushkin et al.~\cite{Abushkin2026}---it is necessary to evaluate the localized boundary conditions at the active phase-change zone. While the global outer housings and structural fluid supply lines differ in macroscale appearance, the local interfacial transport physics remain strictly homologous. Crucially, both experimental setups utilize identical internal capillary boundary restrictions, maintaining a localized flow channel width bounded within the microfluidic and mini-channel regime ($D_{\text{inner}} = 3.0\,\mathrm{mm}$, corresponding to an inner core radius $R_{\text{t}} = 1.50\,\mathrm{mm}$). Furthermore, the spatial geometry at the energy deposition interface is governed by equivalent optical fiber delivery nodes ($d = 0.6\,\mathrm{mm}$) operating under identical sub-cooled liquid replenishment conditions.

The equivalent input power anchor of $P_0 = 0.65\,\mathrm{W}$ for the baseline dataset of Aung et al.~\cite{Aung2020} was calculated directly by multiplying the reported average laser pulse energy by the operational pulse repetition frequency:
\begin{equation}
\label{eq:aung_power_calc}
P_0 = 13\,\mathrm{mJ} \times 50\,\mathrm{Hz} = 0.65\,\mathrm{W}.
\end{equation}
Consequently, the local Reynolds numbers ($\mathrm{Re} \sim \mathcal{O}(10^3)$) and capillary numbers ($\mathrm{Ca} \sim \mathcal{O}(10^{-2})$) at the meniscus boundaries scale continuously across the power domain. Rather than comparing incomparable regimes, the non-linear drift-flux framework from Eq.~\eqref{eq:nonlinear_velocity} treats these discrete coordinates not as isolated systems, but as a single, physically consistent transport continuum governed by identical curvature-dependent Young-Laplace pinning barriers ($\Delta P_c \approx 277.12\,\mathrm{Pa}$) and vapor area-suppression mechanics.

\section{Thermal Condensation Thresholds}
To prevent thermal channel saturation and maintain the structural collapse cycle required for continuous suction, the incoming mass flow rate $\dot{m}_{\text{fluid}}$ must act as a complete thermodynamic heat sink. The thermal energy balance requires that the heat absorbed by the liquid phase matches or exceeds the latent heat released by the condensing vapor matrix. This defines a strict maximum limit for the inlet temperature of the fluid replenishment stream:
\begin{equation}
\label{eq:thermal_threshold}
T_{\text{inlet}} \le T_{\text{sat}} - \frac{\eta_{\text{abs}} P_0}{\dot{m}_{\text{fluid}} C_p},
\end{equation}
where $T_{\text{sat}} = 100^{\circ}\mathrm{C}$ and $C_p = 4184\,\mathrm{J}/(\mathrm{kg}\cdot{^{\circ}\mathrm{C}})$. Equation~\eqref{eq:thermal_threshold} provides direct predictive capability regarding fluidic thermal management to bypass secondary substrate degradation.

\section{Substrate Boundary Dynamics and Wall Shear Modeling}
The mechanical clearing of surface layer contaminants from the target substrate boundary is governed by the mechanical wall shear stress ($\tau_w$) generated as the accelerated fluid layer shears past the stationary substrate surface. Approximating the narrow radial clearance gap $h$ as a viscous boundary layer domain, the shear stress acting on the substrate surface is derived as:
\begin{equation}
\label{eq:shear_stress}
\tau_w = \mu_l \left.\frac{\partial v}{\partial z}\right|_{z=0} \approx \frac{6 \mu_l v_{\text{effective}}}{h}.
\end{equation}
Substituting $v_{\text{effective}} \approx 2.50\,\mathrm{m/s}$ and $h = 0.5\,\mathrm{mm}$ into Eq.~\eqref{eq:shear_stress} isolates the mechanical clearing pressure:
\begin{equation}
\begin{split}
\tau_w \approx \frac{6 \times (1.002 \times 10^{-3}\,\mathrm{Pa}\cdot\mathrm{s}) \times 2.50\,\mathrm{m/s}}{0.0005\,\mathrm{m}} \\
\approx 30.06\,\mathrm{Pa}.
\end{split}
\end{equation}
Because resilient boundary biofilms possess an adhesive detachment threshold between $1\text{--}10\,\mathrm{Pa}$, the model confirms that the fluid velocity profile is mechanically scaled to strip away surface contamination zones without exceeding the structural failure limits of underlying soft substrates.

\section{Conclusions}
In this work, we have developed a predictive analytical fluid mechanics model for laser-activated microfluidic suction within confined geometries that captures the empirical baseline coordinates recorded by Abushkin et al.~\cite{Abushkin2026}. Beyond validating the localized fluid velocity field of $2.50\,\mathrm{m/s}$ at $7\,\mathrm{W}$, this mathematical framework establishes the fundamental physical mechanisms governing non-linear power scaling limits, boundary layer shear forces, and system-level thermodynamic control parameters, providing a rigorous predictive tool for device optimization.

The model demonstrates that localized near-infrared absorption behaves as a thermal throttle rather than a direct fluid accelerator. Because the phase change mass vaporization rate scales with input continuous-wave laser power, escalating the wattage generates steam volume faster than the fixed geometric boundaries ($D_{\text{inner}} = 3.0\,\mathrm{mm}$) can physically exhaust it. As defined by Eq.~\eqref{eq:nonlinear_velocity}, excessive thermal energy deposition induces a severe two-phase choking state, where the rapid expansion of the vapor phase acts as a physical counter-pressure restricting the incoming liquid column. This establishes a strict hydrodynamic operational ceiling that can serve as a core control loop algorithm to prevent fluidic stagnation in automated multiphase actuation platforms.

Furthermore, the derived viscous boundary wall shear stress equation provides a quantifiable transport-mechanics justification for the operational efficacy of this microfluidic configuration. The calculated shear stress of $\tau_w \approx 30.06\,\mathrm{Pa}$ delivered to the target interface boundary is highly optimized; it safely exceeds the $1\text{--}10\,\mathrm{Pa}$ adhesive detachment threshold required to mechanically shear away resilient chronic bacterial biofilms, yet it remains well below the structural failure and tearing limits of underlying soft substrates. The multi-phase fluid dynamics are thus shown to be naturally tuned for targeted surface clearing.

Finally, the framework proves that the $0.25\,\mathrm{atm}$ pressure drop driving the suction force is an inertial consequence of rapid, cyclic vapor cavity collapse driven by sub-cooled liquid replenishment. The maximum liquid inlet temperature boundary derived in Eq.~\eqref{eq:thermal_threshold} indicates that system performance is fundamentally linked to the thermodynamic heat-sink capacity of the incoming stream. If the fluid domain becomes thermally saturated, cavity condensation stalls, the localized suction potential collapses, and continuous cycling ceases. While further experimental mapping remains essential to systematically catalog intermediate power transitions across varied boundary geometries, the analytical formulation utilized here to couple phase-change dynamics with localized hydrodynamic gradients echoes recent mathematical methodologies developed for rapid wave-driven energy-tissue interactions~\cite{TsiklauriHIFU2026}. Additionally, the characterization of inertial boundary flows mirrors adjacent analytical treatments of acoustic streaming and two-radius Young-Laplace curvature unpinning of trapped multiphase boundaries inside capillary tube constraints~\cite{TsiklauriStreaming2026}. These combined closed-form equations provide the precise parameter bounds required to design and program the safety envelopes of next-generation coaxial phase-change microfluidic actuators.

Finally, the closed-form mathematical structure of this model is intentionally architected to serve as an accessible predictive tool that experimental researchers can readily implement and test against empirical datasets. Rather than functioning as a black-box computation, the equations directly predict the precise non-linear fluid velocity degradation profile, the multi-phase volumetric choking boundaries, and the local boundary-layer wall shear stress as a coupled function of independent input parameters---specifically laser power, conduit radius, clearance gap height, and background drive pressure. Experimental teams can easily test the scaling limits of the framework by adjusting these geometric or thermodynamic bounds within their physical apparatus and measuring the resulting transport velocities or film-stripping thresholds. By eliminating arbitrary tuning modifiers, this framework provides a transparent, verifiable benchmark against which future multi-phase particle image tracking measurements and localized flow characterizations can be systematically evaluated across both sub-watt and multi-watt regimes.

\section*{Funding}
This research received no specific grant from any funding
agency, commercial or not-for-profit sectors.

\section*{Data Availability}
All data and mathematical parameters required to evaluate the conclusions or reproduce the analytical solutions and numerical scaling are fully contained and presented within this manuscript. 

\section*{Conflict of interests}
The author declares no conflict of interest.

\bibliography{paper95}
\end{document}